\documentstyle[12pt,fleqn,epsf,epsfig]{article}

\begin{document}

\title{Prevalence of Milnor Attractors and Chaotic Itinerancy
in 'High'-dimensional Dynamical Systems}

\author{Kunihiko Kaneko\\
{\small \sl Department of Pure and Applied Sciences,
College of Arts and Sciences,}\\
{\small \sl University of Tokyo,}\\
{\small \sl Komaba, Meguro-ku, Tokyo 153, Japan}\\
}

\date{}

\maketitle
\begin{abstract}
Dominance of Milnor attractors in 
high-dimensional dynamical systems is reviewed, with the use of globally coupled maps.
From numerical simulations, the threshold number of degrees of freedom
for such prevalence of Milnor attractors is suggested to be $5 \sim 10$, which
is also estimated from an argument of combinatorial explosion of basin boundaries.
Chaotic itinerancy is revisited from the viewpoint of Milnor attractors.  
Relevance to neural networks is discussed.
\end{abstract}

\section{Introduction}

High-dimensional dynamical systems often have many attractors. Indeed,
such systems with multiple attractors have been discussed to be  relevant
to biological memory, information processing,
and differentiation of cell types\cite{KK-Tsuda,Rel}.
There, to make connection with problems in biological networks,
stability of attractors against external perturbation,
switching among attractors by external operation, and spontaneous
itinerancy over several lower dimensional states are important.
In the present paper, we survey a universal aspect with regards to
these points.

One of the simplest model for such high-dimensional
dynamical systems is globally coupled dynamical systems.
In particular,
``globally coupled map" (GCM) consisting of chaotic elements \cite{GCM}
has been extensively studied, as a simple prototype model.
A standard  model for such GCM is given by

\begin{equation}
x_{n+1}(i)=(1-\epsilon )f(x_{n}(i))+\frac{\epsilon }{N}\sum_{j=1}^N f(x_{n}(j))
\end{equation}
where $n$ is a discrete time step and $i$ is the index of an
element ($i= 1,2, \cdots ,N$ = system size), and $f(x)=1-ax^{2}$.
The model is just a mean-field-theory-type extension of coupled map
lattices (CML)\cite{CML}.  

Through the average interaction, elements are tended to oscillate synchronously,
while chaotic instability leads to destruction of the coherence.
When the former tendency wins, all elements oscillate coherently,
while elements are completely desynchronized in the limit of
strong chaotic instability.  Between these cases, elements split
into clusters in which they oscillate coherently.
Here a cluster is defined
as a set of elements in which $x(i)=x(j)$.
Attractors in GCM are classified by the number of synchronized
clusters $k$ and the number of elements for each cluster $N_i $.
Each attractor is coded by the
clustering condition $[k,(N_1 ,N_2 ,\cdots,N_k )]$.

As has been studied extensively,
the following phases appear successively with the increase of
nonlinearity in the system ($a$ in the above logistic map case)
\cite{GCM}:

(i) {\bf Coherent phase};
(ii) {\bf Ordered phase}; 
(iii) {\bf Partially ordered phase};
(iv) {\bf Desynchronized  phase}:  

In (i), a completely synchronized  attractor ($k=1$) exists, while
all attractors consist of few ($k=o(N)$) clusters in (ii).
In (iii), attractors with a variety of clusterings
coexist, while most of them have many clusters ($k=O(N)$).
Elements are completely desynchronized, i.e., $k=N$ for all 
(typically single) attractors in (iv).
The above clustering behaviors
have universally been confirmed in a variety of systems (see also \cite{Yuri-book}).

In the partially ordered (PO) phase,
there are a variety of attractors with a different number of clusters,
and a different way of partitions $[N_1,N_2, \cdots ,N_k]$.
In this phase, there are a variety of partitions as attractors.
As an example, we measured the fluctuation of the partitions,
using the probability $Y$ that two elements fall on the same cluster.
In the PO phase, this $Y$ value varies by attractors,
and furthermore, the variation remains finite even in the limit of $N \rightarrow \infty$
\cite{GCM-part,Vul}.  In other words, there is no 'typical' attractor
in the thermodynamic limit.  This is similar with the
'non-self-averaging' in Sherrington-Kirkpatrick model in spin glass\cite{SG}.

\section{Attractor Strength and Milnor Attractors}

In the partially ordered (PO) phase and also in some part of ordered phase,
there coexist a variety of attractors depending on the partition.
To study the stability of an attractor against perturbation, we introduce the
return probability $P(\sigma)$, defined as follows\cite{GCM-Milnor,KK-Milnor}:
Take an orbit point $\{x(i)\}$ of an attractor in an
$N$-dimensional phase space,
and perturb the point to $x(i)+\frac{\sigma}{2}rnd_i$, where
$rnd_i$ is a random number taken from $[-1,1]$, uncorrelated for all elements
$i$. Check if this perturbed point returns to the original attractor
via the original deterministic dynamics (1).
By sampling over random perturbations and the time of the application of
perturbation, the return probability $P(\sigma)$ is defined as
(\# of returns )$/$ (\# of perturbation trials).
As a simple index for robustness of an attractor, it is
useful to define $\sigma_c$ as the largest $\sigma$ such that $P(\sigma)=1$.
This index measures what we call the {\sl strength} of an attractor.

The strength $\sigma_c$ gives a minimum distance between
the orbit of an attractor and its basin boundary.
Note that $\sigma_c$ can be small,
even if the basin volume is large, if the attractor is located near the basin
boundary.

In contrast with our naive expectation from the concept of
an attractor, we have often observed
`attractors' with $\sigma_c =0$, i.e.,
$P(+0) \equiv \lim_{\delta \rightarrow 0}P(\delta) <1$.
If $\sigma_c = 0$ holds for a given state, it cannot be an
``attractor" in the sense with asymptotic stability,
since some tiny perturbations kick the orbit out of the ``attractor".
The attractors with $\sigma_c =0$
are called Milnor attractors\cite{Milnor0,Milnor}.  In other words,
Milnor attractor is defined as an attractor that is unstable
by some perturbations of arbitrarily small size, but globally attracts
orbital points.  (Originally, Milnor proposed to include
all states with the basin of attraction of a positive Lebesgue measure,
into the definition of an attractor\cite{Milnor0}.
Accordingly,  attractors by his definition include also the usual attractor with
asymptotic stability.  Here we call Milnor attractor, only if it does not
belong to the latter.  If this Milnor attractor is chaotic,
the basin is considered to be riddled \cite{riddle,riddle-CM}.  This is the
case for the present GCM model.)
Since it is not asymptotically
stable, one might, at first sight, think that it is
rather special, and appears only at a critical point like the
crisis in the logistic map\cite{Milnor0}.  To our surprise, 
the Milnor attractors are rather commonly observed around the border between
the ordered and partially ordered  phases in our GCM (see Fig.2).
Attractors with $\sigma _c=0$ often have a large basin volume, which
sometimes occupy almost 100 \% of the total phase space.

\begin{figure}
\epsfig{file=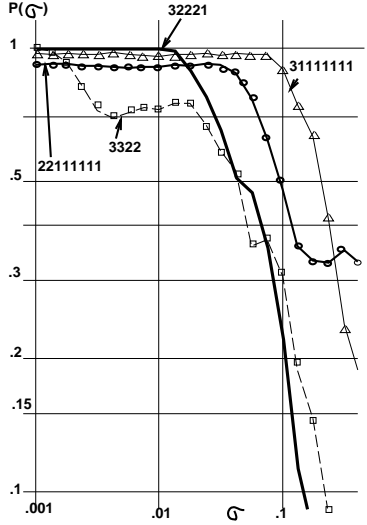,width=.35\textwidth}
\caption{
$P(\sigma)$ for 4 attractors
for $a=1.64$, and $N=10$.  
10000 initial conditions are randomly chosen, to make samplings.
$P(\sigma)$ is estimated by sampling over 1000 possible perturbations
for each $\sigma$.
Plotted are robust attractors [32221] ($\sigma_c \approx .01$), [3322] ($\sigma_c \approx $.0012),
and Milnor attractors [31111111], [22111111]. The basin volume of the
latter two occupies 42\% and 29\% of the phase space.
}
\end{figure}

\begin{figure}
\epsfig{file=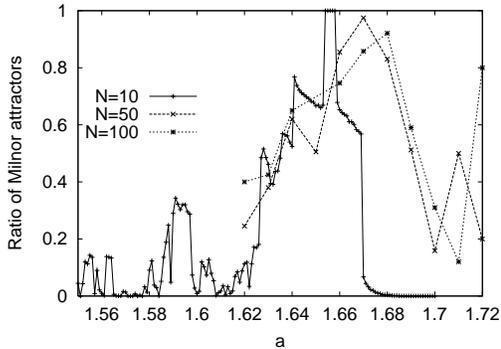,width=.5\textwidth}
\caption{
The fraction of the basin ratio of Milnor attractors, 
plotted as a function of $a$, for $N=10,50,100$
}
\end{figure}

With regards to the basin volume of Milnor attractors,
the change of the behavior of the GCM (1) 
with the increase of the parameter $a$ is summarized as follows:
{\sl a few attractors with small numbers of clusters;}
$\rightarrow$ {\sl increase of the number of attractors with
stable and Milnor attractors coexisting;}
$\rightarrow$ {\sl decrease of the number of attractors with some remaining
Milnor attractors with large basin fractions;}
$\rightarrow$ {\sl only a single or a few stable  attractors
with complete de-synchronization
remain. Milnor attractors no longer exist.}

Then, why is the basin volume of Milnor attractors so large for some parameter regimes?
To answer the question, robustness of global attraction is a key.
Note there are a large number of attractors
at the border between O and PO.  Most of the attractors  lose the stability around the
parameter regime successively.  When the stability of an attractor
is lost, there appears a set of points in
the vicinity of the attractor, that are kicked out of it through the temporal
evolution, while
the global attraction still remains.  This is a reason why fragile attractors
are dominant around the PO phase.  In Fig.2, we have plotted the sum of
basin volume rates for all the Milnor attractors. Dominance of
Milnor (fragile) attractors is clearly seen.

Attractors are often near the crisis point
and lose or gain the stability at many parameter values in the PO phase.
Furthermore, the stability of
an attractor often shows sensitive dependence on the parameter.
It is interesting to see how $P(+0)$ and basin volume change with the
parameter $a$,
when an attractor loses asymptotic stability.  As shown in \cite{correction},
the basin volume of an attractor often has a peak when the
it loses the stability and then decreases slowly as $P(+0)$ gets
smaller than unity, where the attractor becomes a Milnor one.  Although the
local attraction
gets weaker as $P(+0)$ is smaller than 1, the global attraction remains.
It is also noted that
if $P(+0)$ equals 1 or not often sensitively depends on the parameter $a$, 
while the basin volume shows smooth dependence on the parameter.
The basin volume reflects on global attraction, while
the $P(+0)$ depends on local structure in the phase space
with regards to collision of an attractor and its basin boundary.
This is the reason why the former has a smooth dependence of the parameter,
in contrast to the latter.

{\sl remark}

Coexistence of attractors with different degrees of stability
makes us expect that noise is relevant  to the choice
of the attractor the GCM settles to. 
One might then suspect that such Milnor attractors must be weak
against noise.  Indeed, by a very weak noise with the amplitude $\sigma$,
an orbit at a Milnor attractor is kicked away, and if the orbit is reached 
to one of attractors with $\sigma_c >\sigma$, it never comes back to the Milnor attractor.
In spite of this instability, however, an orbit kicked out from a Milnor attractor 
is often found to stay in the vicinity of
it, under a relatively large noise\cite{KK-Milnor}. 
The orbit comes back to the original Milnor attractor
before it is kicked away to other attractors with $\sigma_c >\sigma$.
Furthermore, by a larger noise, orbits sometimes are more attracted 
to Milnor attractors.  Such attraction is possible, since Milnor attractors
here have global attraction in the phase space, in spite of their local
instability.

\section{Magic Number $7 \pm 2$ in Dynamical Systems?}

Milnor attractors can exist in low dimensional
dynamical systems like a two-dimensional map as well.
When changing the parameter of a dynamical system, the basin boundary of an
attractor may move until, for a specific value of the parameter, the
basin boundary touches the attractor.
Then,  if the attractor has
a positive measure of initial conditions forming the basin of attraction,
it becomes a Milnor attractor.  Generally speaking, however,
the above situation occurs only for very specific parameter values,
and it is not naively expected that the Milnor attractors exist
with a positive measure in the parameter space.

However, as was shown in the last section,
Milnor attractors are found to be rather prevalent, occurring not only for specific
isolated parameter values.
Such dominance of Milnor attractors is often found in high-dimensional
dynamical systems, for example coupled maps with 10
degrees of freedom or so.
Then, the question we address now is why
can there be so many Milnor attractors in a ``high-dimensional''
dynamical system, and what number of degrees of freedom
is sufficient for constituting such `high' dimensionality.

We computed the average
basin fraction of Milnor attractors
over the parameter interval $1.55<a<1.72$.  In Fig.3, this fraction is plotted
as a function of the number of degrees of freedom $N$.
The increase of the average basin fraction of Milnor attractors with $N$ is clearly visible
for $N \approx (5 \sim 10)$, while it levels off for
$N>10$.  Indeed, such increase  of Milnor attractors with the
degrees of freedom $5\sim10$ seems to be universal
in a partially ordered phase in globally coupled chaotic system\cite{KK-magic}.
(Here, the degree of freedoms we use in the present paper is the number of units that has
orbital instability. For example, if we choose a coupled system of
$N$ Lorenz equations, the degrees we mention is not $3N$, but $N$.)

\begin{figure}
\epsfig{file=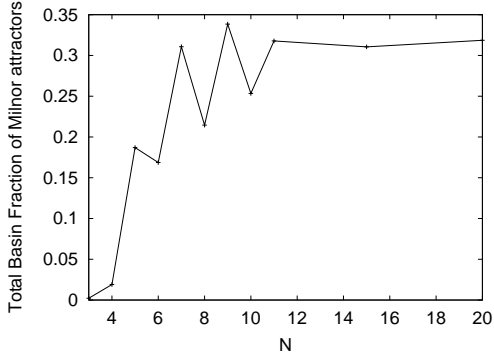,width=.5\textwidth}
\caption{
The average fraction of the basin ratio of Milnor attractors.
After the basin fraction of Milnor attractor is computed
as in Fig.1, the average of the ratios for parameter values
$a=1.550,1.552,1.554, \cdots 1.72$ is taken.
This average fraction is plotted as a function of $N$.
(In this class of models, the fraction of Milnor attractors
is larger for odd $N$ than for even $N$.  Note that  two clusters with
equal cluster numbers and anti-phase oscillations generally have less chaotic instability.  
A globally coupled map with an even number of elements
allows for equal partition into two clusters.  This gives a plausible
explanation for the smaller instability for even $N$ system.)
}
\end{figure}

Now we discuss a possible reason how the dominance of Milnor attractors appears.
In a system with identical elements, due to the symmetry, there are at least
\begin{math}
M(N_1,N_2,\cdots ,N_k)=\frac{N!}{\prod_{i=1}^{k} N_i !}
\prod_{\mathrm{over sets of} N_i=N_j}\frac{1}{m_{\ell}!}
\end{math}
attractors for each clustering condition, where $m_{\ell}$ is the
number of clusters with the same value $N_j$.
Then, a combinatorial explosion in the number of attractors can be expected
when many of the clustering conditions are allowed as attractors.
For example, the permutation of $N$ elements leads to $(N-1)!$
possibilities and one might expect the number of  attractors
to be of this order.  On the other hand, the phase space volume in a coupled
system expands only exponentially with $N$.  Typically the combinatorial explosion
outruns the exponential increase around $N \approx (5 \sim 10)$.
(For example compare $2^N$ and $(N-1)!$.  The latter surpasses the former
at $N=6$.)  Hence, the attractors crowd\cite{Wiesen} in the phase space 
and the stability of each attractor may be lost.

However, this argument seems to be incorrect.  
We have computed the number of attractors and compared with the basin fraction
of Milnor attractors.  As has been shown\cite{KK-magic},
the dominance of the Milnor attractors is not necessarily
observed when the number of attractors is high.
With increasing $a$, the fraction remains large even when the number of attractors
has already decreased substantially.
Since the basin volume of each of the attractors is far from being equal,
the explosion in the number of attractors does not necessarily mean that
the basin volume for each and every attractor should be very small.
Indeed, according to our numerical results, for the parameter region where
the Milnor attractors dominate, the number of Milnor attractors is not
so high and the basin fraction of only a few Milnor attractors occupies
almost all of phase space.

In the above sequence, the dominance of the Milnor attractors is observed
when many attractors have disappeared.  Therefore, we can revise the
first explanation on the dominance by replacing the combinatorial explosion
in the number of attractors themselves by the combinatorial explosion
in the number of basin boundaries that separate the  attractors.  
For the parameter region where many attractors
start to disappear, there remain basin  boundary points separating such 
(collapsed) attractors and the remaining attractors.  

Now, we need to discuss how the 
distance between an attractor and its basin boundary changes with $N$.
Consider a one-dimensional phase space, and a 
basin boundary that separates the regions of $x(1)>x^*$ and $x(1)<x^*$,
while the attractor in concern exists at around $x(1)=x_A<x^*$, and the
neighboring one at around  $x(1)=x_B>x^*$.
Now consider a region of $N$-dimensional phase space $x_A<x(i)<x_B$.
If the region is partitioned by (basin) boundaries at $x(i)=x^*$ for $i=1,\cdots N$,
it is partitioned into $2^N$ units.  Since this partition is just 
a direct product of the original partition by $x(1)=x^*$,
the distance between each attractor and the basin boundary
does not change with $N$. (For example, consider the extreme case that
$N$ identical maps are uncoupled ($\epsilon=0$).)

On the other hand, consider a boundary given by some condition for 
$(x(1),\cdots , x(N))$, represented by a (possibly very complex)
hyperplane $C(x(1),\cdots , x(N))=0$.
In the present system with global (all-to-all) couplings,
many of the permutational changes of $x(i)$ in the condition 
also give basin boundaries.  Generally, the condition for the basin
can also have 
clustering $(N_1,\cdots ,N_k)$, since the attractors are clustered 
as such.  Then the condition obtained by the 
permutation of $C(x(1),\cdots , x(N))=0$ gives a basin boundary also
(or one can say that $C(x(1),\cdots , x(N))=0$ itself satisfies
such permutational symmetry).
Then, the basin boundary has
$M(N_1,\cdots ,N_k)$ segments transformed into each other  by the permutations.
The number of such segments of the boundaries   
increases combinatorially with $N$.  Roughly speaking, the sum of $M(N_1,\cdots,N_k)$
 increases in the order of $(N-1)!$, when 
a variety of clusterings is allowed for the boundary.  Now the $N$-dimensional
phase space region is partitioned by $O(N-1)!)$ basin boundary segments.
Recalling that the distance between an attractor and the basin boundary
remains at the same order for the partition of the order of $2^N$, 
the distance should decrease if $(N-1)!$ is larger than $2^N$.  Since 
for $N>5$, the
former increases drastically faster than the latter, 
the distance should decrease drastically for $N>5$.  Then 
for $N>5$, the probability that a basin boundary touches with an attractor itself
will be increased.
Since this argument is applied for any attractors and their
basin boundary characterized by complex 
clusterings having combinatorially large $M(N_1,\cdots ,N_k)$,
the probability that an attractor touches its basin boundary
is drastically amplified for $N>5$. 
Although this explanation may be rather
rough, it gives a hint to why Milnor attractors
are so dominant for $N \stackrel{>}{\approx} (5 \sim 10)$.
We surmise that this is the reason why
Milnor attractors are dominant in our model 
at $(1.64 \sim 1.67)$ when $N \stackrel{>}{\approx}(5 \sim 10)$.

Since the above discussion is based mainly on simple combinatorial
arguments, it is expected that the dominance of Milnor attractors
for $N \stackrel{>}{\approx}(5 \sim 10)$ may be rather common at some
parameter region in 
high dimensional dynamical systems.  It is interesting that
pulse-coupled oscillators with global coupling also show
the prevalence of Milnor attractors for $N \geq 5$
\cite{pulse}.

One might expect that permutational symmetry us necessary
for the prevalence of Milnor attractors.
For example, in the GCM (1),
the permutation symmetry arising from identical elements leads to a
combinatorial explosion in the number of attractors.
Then, one may wonder whether the prevalence of Milnor attractors
is possible only for such highly symmetric systems.
We have therefore studied a GCM with heterogeneous parameters.
Although the fraction seems to be  smaller than in the homogeneous case,
Milnor attractors are again observed and 
their basin volume is rather large for some parameter region.
As in the symmetric case, the basin fraction of Milnor attractors
increases around $N \approx (5 \sim 10)$.

Note that even though complete synchronization between two elements is lost,
clusterings as with regards to the phase relationships can exist.
(As for such phase synchronization of chaotic elements see
\cite{Pikovsky0}).
Indeed, there are two groups when considering the oscillations of phases
as large-small-large... and small-large-small..., that
are preserved in time for many attractors. Furthermore,
finer preserved phase relationships can also exist.  Similarly, it is natural to expect
an explosion in the number of the basin boundary points for some parameter regime.
Accordingly the argument
on the dominance of Milnor attractors for a homogeneous GCM
can be applied here to some degree as well.

It is also interesting to note that in
Hamiltonian dynamical systems, agreement with thermodynamic behavior is often
observed only for degrees of freedom higher than $5 \sim 10$\cite{Sasa}.
Considering the combinatorial complexity woven by  all the possible Arnold webs
(that hence may be termed ``Arnold spaghetti''), the entire phase space volume
that expands only exponentially with the number of degrees of freedom
may be covered by webs, resulting in
uniformly chaotic behavior.  If this argument holds, the
degrees of freedom required for thermodynamic behavior can also be
discussed in a similar manner.

\section{Chaotic Itinerancy}

In the PO phase,  orbits often make itinerancy over several
ordered states with partial synchronization of elements, through  highly chaotic states.
This dynamics, called chaotic itinerancy (CI),
is a novel universal class in high-dimensional dynamical systems\cite{special}.
In the CI, an orbit successively itinerates over
such ``attractor-ruins", ordered motion with some coherence among elements.
The motion at ``attractor-ruins" is quasi-stationary.
For example, if the effective degrees of freedom is two,
the elements split into two groups, in each of which elements
oscillate almost coherently.  The system is in the vicinity of a
two-clustered state, which, however, is not a stable attractor, but keeps 
attraction to its vicinity globally within the phase space.
After staying at an attractor-ruin, an orbit eventually exits from it.
This exit arises from orbital instability.  In the above example, the synchronization
among the two groups is increased.
Then, as is straightforwardly seen in the model equation 
(1), the dynamics are approximately given by $x_{n+1}=f(x_n)$, which
has stronger orbital instability than a clustered state.  With this
instability the state enters into a high-dimensional chaotic
motion without clear coherence.  (Here it is interesting to note that
the effective degrees of freedom decreases before it goes to a high-dimensional state).
This high-dimensional state
is again quasi-stationary, although there are some holes connecting
to the attractor-ruins from it.  Once the orbit is trapped at a hole, it
is suddenly attracted  to one of attractor ruins, i.e., ordered states with low-dimensional
dynamics.

This CI dynamics
has independently been found in a model of neural dynamics by Tsuda \cite{Tsuda},
optical turbulence \cite{Ikeda}, and in GCM\cite{GCM,GCM-CC}.
It provides an example
of successive changes of relationships among elements.

There seem to be several types of ``chaotic itinerancy"
covered by this general definition for it.
It can roughly be classified according to the degree of
correlation between the ordered states visited
successively.  The correlation is high if the paths for the transitions between the ordered
states are narrow, and the probabilities for visiting the next ordered
state are rather low.  On the other hand, the
correlation is low when  the memory of the previous sate is lost due to
high-dimensional chaos during the transition.

Still, the systems with chaotic itinerancy studied so far commonly have
a small number of positive Lyapunov exponents and many exponents close to zero.
As a result, the dimension of the global attractor is high, while
the path in the phase space is restricted.

Note that the Milnor attractors satisfy
the condition of the above ordered states constituting chaotic itinerancy.
Some Milnor attractors we have found keep
global attraction, which is consistent with the observation
that the attraction to ordered states in chaotic itinerancy
occurs globally from a high-dimensional chaotic state.
Attraction of an orbit to precisely a given attractor requires infinite time, and before 
the orbit is really settled to a given
Milnor attractor, it may be kicked away.
Then, the long-term dynamics can be constructed as
the successive alternations to the attraction to, and
escapes from, Milnor attractors.  
If the attraction to robust attractors from a given Milnor attractor is
not possible, the long-term dynamics with the noise strength $\rightarrow +0$ is
represented by successive transitions over Milnor attractors.
Then the dynamics is represented by transition matrix over among Milnor attractors.
This matrix is generally asymmetric: often, there is a
connection from a Milnor attractor A to a Milnor attractor B, but
not from B to A.  The total dynamics is represented by the motion over
a network, given by a set of directed graphs over Milnor attractors.
In general, the `ordered states' in CI may not be exactly Milnor attractors but can be
weakly destabilized states from Milnor attractors.  Still,
the attribution of CI to Milnor attractor network dynamics is expected to
work as one ideal limit.

{\sl remark}

Computability of chaotic itinerancy has a serious problem, since
switching process over Milnor attractor network in the noiseless case
may differ from that of the case with the limit of noise $\rightarrow +0$,
or from that obtained by a digital computer with a finite precision.
For example, once the digits of two variable $x(i)=x(j)$ 
agree down to the lowest bit, the values never split again,
even though the state with the synchronization of the two elements may be
unstable\cite{GCM94}.  As long as digital computation is adopted,
it is always possible that an orbit is trapped to such unstable state.
(See \cite{Pikovsky}, for one technique to resolve this numerical problem).

In each event of switching, which Milnor attractor is visited next
after the departure from a Milnor attractor may depend on the precision, or
on any small amount of noise.
Here it may be interesting to note
that there are similar statistical features between (Milnor attractor)
dynamics with a riddled basin
and undecidable dynamics of a universal Turing-machine\cite{Saito-KK}.

\section{Relevance to Neural Networks}

When one considers (static) memory in terms of dynamical systems,
it is often adopted that each memory is assigned into an attractor.
Here, a system with many attractors is desirable as such system.  Then, 
existence of Milnor attractors may lead us to suspect the
correspondence between a (robust) attractor and memory.

Here, it may be interesting to recall
that the  term magic number $7 \pm 2$ was originally coined in psychology
\cite{Miller}. It was found that
the number of chunks (items) that is memorized in short term memory
is limited to $7 \pm 2$.  Indeed with this number $7 \pm 2$,
the fraction of basins for Milnor attractors
increases.  Since possible explanation is based only on
combinatorial arguments, this
`magic number $5 \sim 10$' in dynamical systems
does not strongly depend on the choice of specific models.
Then, it may be interesting to
discuss a possible connection of it with the original magic number $7 \pm 2$ in psychology.
(see also \cite{TsudaNicolis} for a pioneering
approach to this problem from a viewpoint of chaotic dynamics).
To memorize $k$ chunks of information including their order (e.g., a phone number of $k$ digits)
within a dynamical system,
it is natural to assign each memorized state to an
attractor of a $k$-dimensional dynamical system
(unless rather elaborate mechanisms are assumed).
In this $k$ dimensional phase space, a combinatorial variety of attractors
has to be presumed in order to assure a sufficient variety of memories.
Then, if our argument
so far is applied to the system, Milnor attractors may be dominant
for $k > (5 \sim 10)$.
If this is the case, the state represented by a Milnor attractor may be
kicked out by tiny perturbations.  Thus robust
memory may not be possible for information that contains more than $7 \pm 2$ chunks.
Possibly, this argument can also be applied to
other systems that adopt attractors as memory,
including most neural networks.

(Of course, the present argument should mainly be applied to systems with
all-to-all couplings, or to highly connected network systems.
If the connections are hierarchically ordered, the number of memory items can be
increased.  The often adopted module structure is relevant for this purpose.)

This argument does not necessarily imply that Milnor attractors are irrelevant to
cognitive processes.  For a dynamical system
to work as a memory, some mechanism to write down and read it out is necessary.
If the memory is given in a robust attractor, its information processing
is not so easy, instead of its stability.  Milnor attractors may provide 
dynamic memory \cite{Tsuda,KK-Tsuda} allowing for
interface between outside and inside, external inputs and internal
representation.
In a Milnor attractor, 
some structure is preserved, while it
is dynamically connected with different attractors.
Also, it can be switched to different memory by any small inputs.
The connection to other attractors is neither one-to-one nor random.  It is
highly structured with some constraints.

Searches with chaos itinerating
over attractor ruins has been discussed in\cite{Freeman,Tsuda} with
a support in an experiment on the olfactory bulb\cite{Freeman}.
Freeman, through his experiments,
proposed that the chaotic dynamics
corresponds to a searching state for a variety of memories, represented by
attractors \cite{Freeman}, while evidence from human scalp EEG showing
chaotic itinerancy is also suggested\cite{Freeman2}.

We note that the Milnor attractors in our GCM model provide a candidate
for such a searching state, because of connection to a variety of
stronger attractors which possibly play the role of rigidly memorized states.
Stability of Milnor attractors by some noisy inputs also
supports this correspondence.

{\sl acknowledgments}

The work is partially supported by Grant-in-Aids for Scientific
Research from the Ministry of Education, Science, and Culture of Japan
(11CE2006).


\begin{thebibliography}{999}
\bibitem{correction}
See Fig.10 of ref. \cite{KK-Milnor}, where the right axis corresponding to the
thick line should be read as $P(+0)$, not $\sigma_c$.  Also the value $0.5$ at the
right axis of the right figure should be 1.0.
\end{thebibliography}
\end{document}